\documentclass[12pt,twocolumn]{iopart}
\usepackage{graphicx}% Include figure files

\usepackage{iopams}

\newcommand{\ket}[1]{\vert #1 \rangle}

\newcommand{\meanvalue}[3]{\langle #1 \vert #2 \vert #3 \rangle}
\newcommand{\ketbra}[2]{\vert #1 \rangle \langle #2 \vert}

\begin{document}

\title[Tripartite entanglement transfer in a CQED open system]{Tripartite quantum state mapping and discontinuous entanglement transfer in a cavity QED open system}

\author{M. Bina$^{1,2}$, F. Casagrande$^{1,2}$, M. G. Genoni$^{1,2}$, A. Lulli$^{1,2}$ and M. G. A. Paris$^{1,2,3}$}

\address{$^1$ Dipartimento di Fisica, Universit\`{a} di Milano, I-20133 Milano, Italy}
\address{$^2$ CNISM, UdR Milano, I-20133, Milano, Italy}
\address{$^3$ ISI Foundation, I-10133, Torino, Italy}
\ead{federico.casagrande@mi.infn.it}

\begin{abstract}
We describe the transfer of quantum information and entanglement
from three flying (radiation) to three localized (atomic) qubits via
cavity modes resonantly coupled to the atoms, in the presence of a
common reservoir. Upon addressing the full dynamics of the resulting
nine-qubit open system, we find that once the cavities are fed,
fidelity and transferred entanglement are optimal, while their peak
values exponentially decrease due to dissipative processes. The
external radiation is then turned off and quantum correlations
oscillate between atomic and cavity qubits. For a class of mixtures
of W and GHZ input states we deal with a discontinuous exchange of
entanglement among the subsystems, facing the still open problem of
entanglement sudden death and birth in a multipartite system.

\end{abstract}

\maketitle

\section{Introduction}
As early as in 1935 Einstein with Podolski and Rosen \cite{EPR} and
Schr\"{o}dinger \cite{Sch} drew the attention on the correlations in
quantum composite systems and the problems raised by their
properties. Much later, theoretical \cite{Bell} and experimental
\cite{Aspect} cornerstones elucidated the issue of nonlocality.
Entanglement is currently viewed as the key resource for quantum
information processing \cite{Nielsen}, where it allowed a number of
achievements such as teleportation \cite{Bennett}, cryptography
\cite{Gisin} and enhanced measurements \cite{entame}. The deep
meaning of multipartite entanglement, its quantification and
detection \cite{Guhne} as well as its possible applications, are the
object of massive investigation touching the heart of quantum
physics and a great variety of its branches. In this paper we study
the entanglement dynamics of a multipartite open system in cavity
quantum electrodynamics (CQED) \cite{Haroche_Raimond} by a model
that is not so far from physical implementation with current
technology of optical cavities \cite{Nussmann}. In particular we
study a paradigmatic example to investigate multipartite
entanglement transfer and swapping \cite{Paternostro,Casagrande},
which are fundamental processes for the implementation of quantum
interfaces and memories for quantum networks \cite{Cirac, Ser06}. In
addition we deal with the recently discovered \cite{YuPRL} and
observed \cite{Almeida} phenomenon of entanglement sudden death
(ESD) (and birth (ESB)), consisting in an abrupt vanishing (and
raising) of quantum correlations. The problem is nowadays still open
with respect to the multipartite entanglement classification, but we
can shed some light at least on the ESD and ESB of the fully
tripartite entanglement. The interplay among all these aspects in
the presence of external environments is investigated.

\section{The multipartite open system model}
We consider three entangled radiation modes which are injected and
resonantly coupled with  three separated optical cavities, each of
them containing a trapped two-level atom whose transition frequency
is resonant with the cavity mode. Each cavity mode is coupled to an
external environment by an amplitude damping channel with a decay
rate $k$ and each atom can spontaneously emit a photon with a decay
$\gamma$. The evolution of the whole system density operator
$\hat{\rho}(t)$ can be described by a Master Equation (ME) in the
Lindblad form that we have solved by means of the Monte Carlo Wave
Function method \cite{MCWF}. We identify a set of six collapse
operators $\hat{C}_{c,J}=\sqrt{\tilde{k}}\hat{c}_J$ and
$\hat{C}_{a,J}=\sqrt{\tilde{\gamma}}\hat{\sigma}_J $ ($J$=A,B,C),
where $\tilde{k}=k/g_a$ and $\tilde{\gamma}=\gamma/g_a$ are
dimensionless cavity and atomic decay rates scaled to the
atom-cavity coupling constant $g_a$ (taken equal for each cavity
mode-atom subsystem). The effective Hamiltonian is $\label{H_eff}
\hat{\mathcal{H}}_e=\frac{\hat{\mathcal{H}}^I}{g_a}-\frac{i\hbar}{2}\sum_{J}\left
[
\hat{C}^{\dag}_{c,J}\hat{C}_{c,J}+\hat{C}^{\dag}_{a,J}\hat{C}_{a,J}\right
]$, with the system Hamiltonian in the interaction picture
$\label{H_int} \hat{\mathcal{H}}^I=\sum_{J}\big[\hbar
g_a(\hat{c}_J\hat{\sigma}^{\dag}_J+\hat{c}^{\dag}_J\hat{\sigma}_J)+i\hbar
g_c(\hat{c}_J\hat{f}^{\dag}_J-\hat{c}^{\dag}_J\hat{f}_J)\big]$. Here
$\hat{c}_J,\hat{c}^{\dag}_J$ ($\hat{f}_J,\hat{f}_J^{\dag}$) are the
quantum harmonic oscillator operators for the cavity (external
radiation) modes, $\hat{\sigma}_J,\hat{\sigma}_J^{\dag}$ the raising
and lowering operators for the atomic qubits, and $g_c$ the
cavity-input
field coupling constant.\\
\subsection{Subsystems dynamics, state mapping and entanglement transfer}
\begin{figure}[h]
\includegraphics[width=0.23\textwidth]{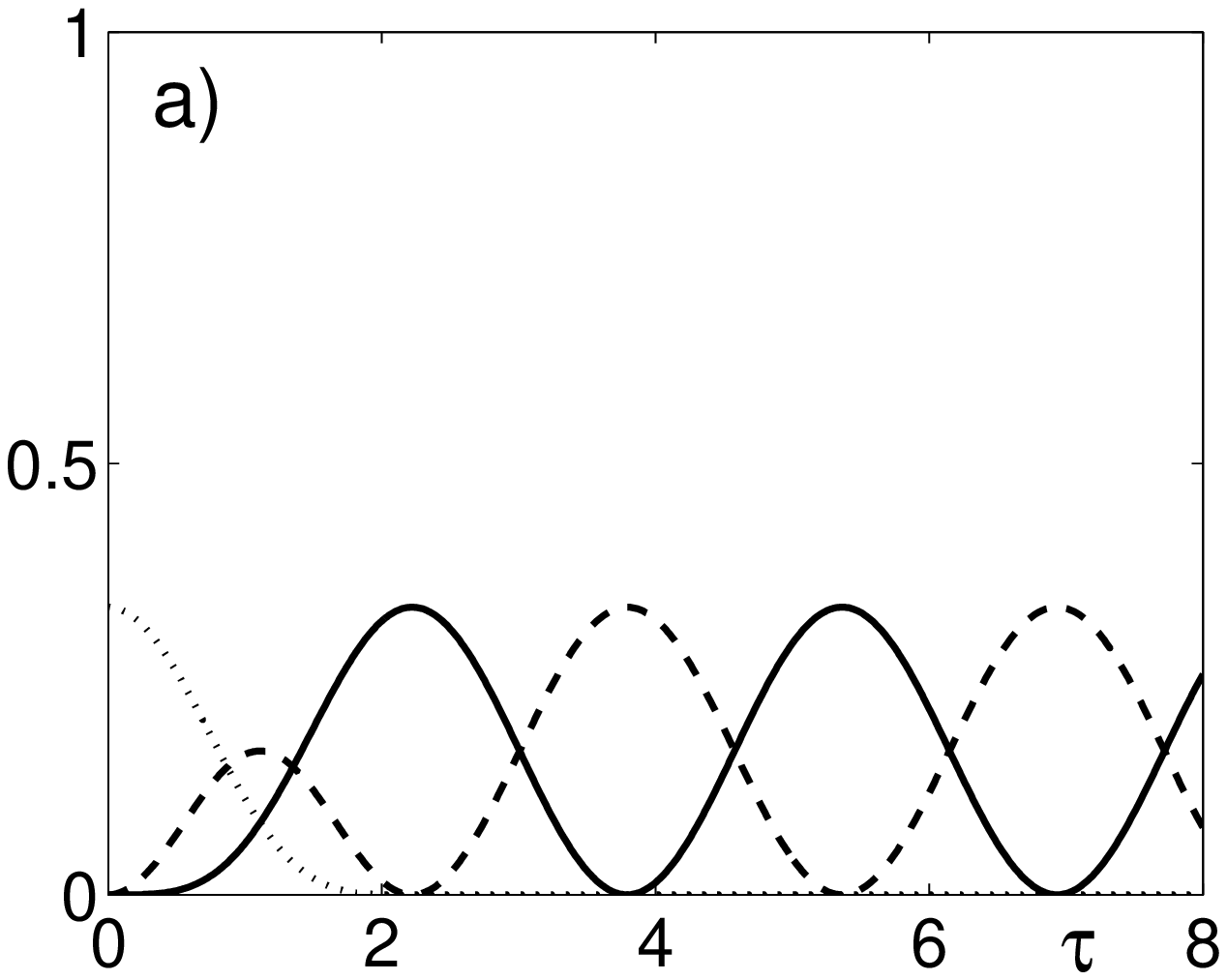}
\includegraphics[width=0.23\textwidth]{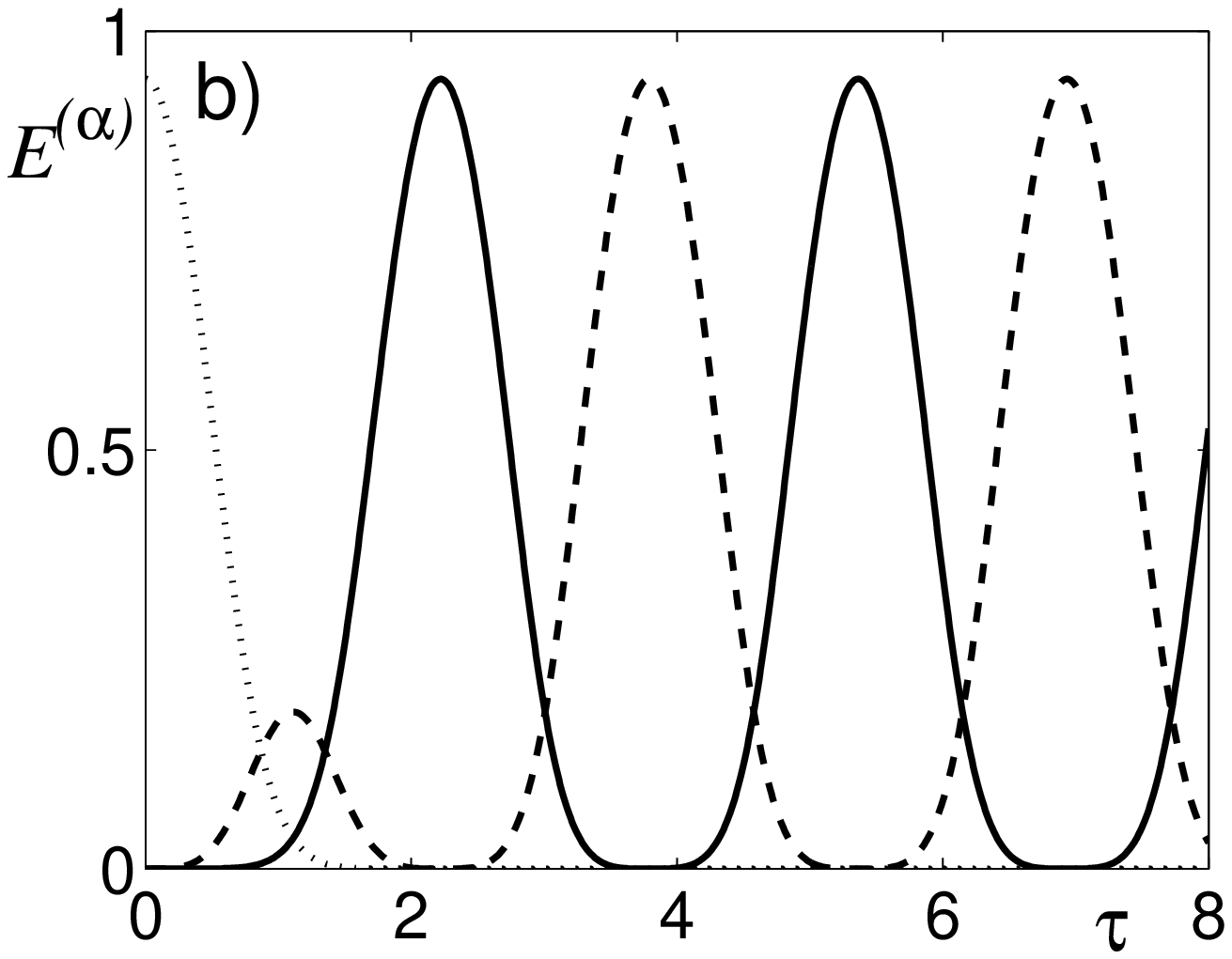}
\includegraphics[width=0.23\textwidth]{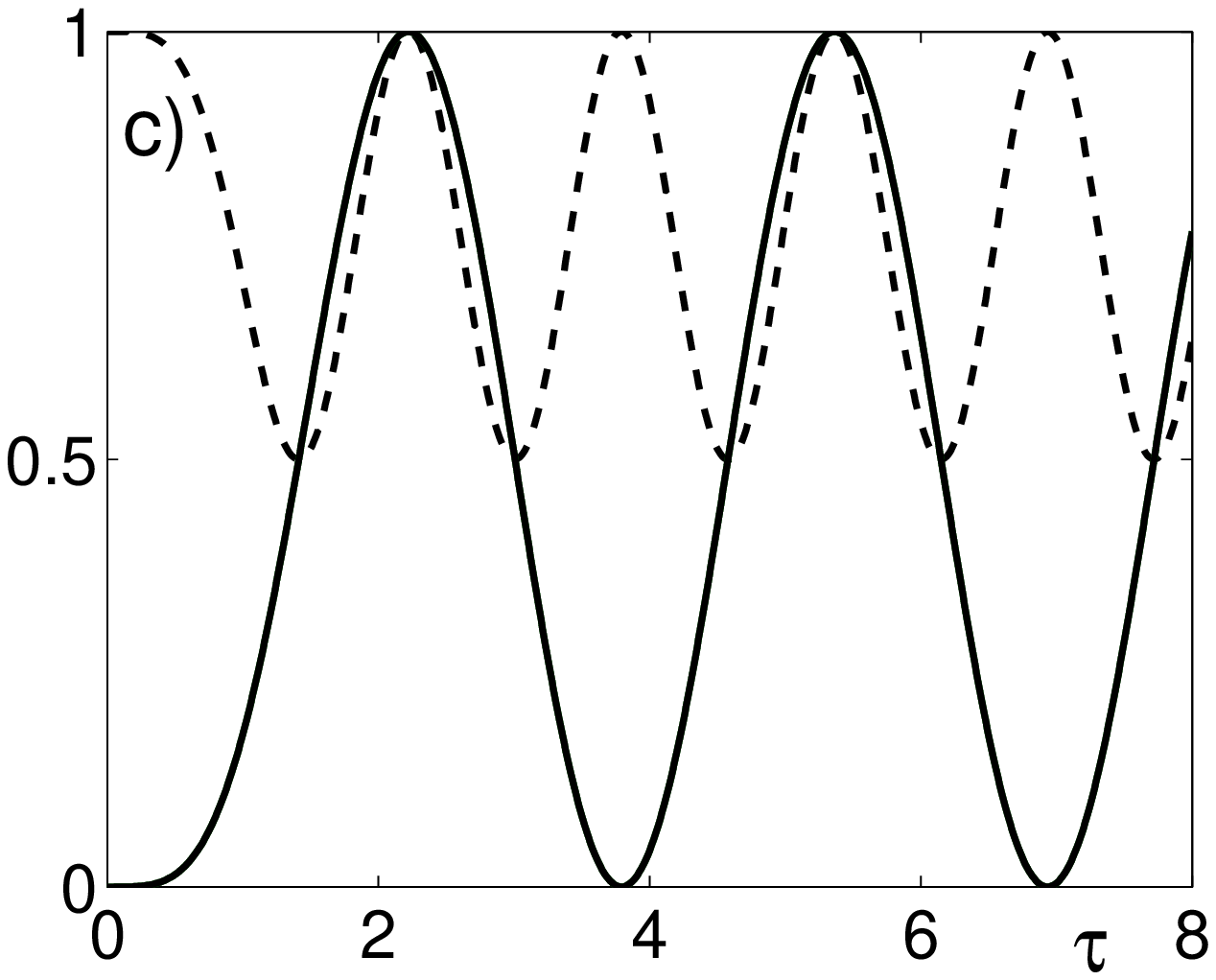}
\includegraphics[width=0.23\textwidth]{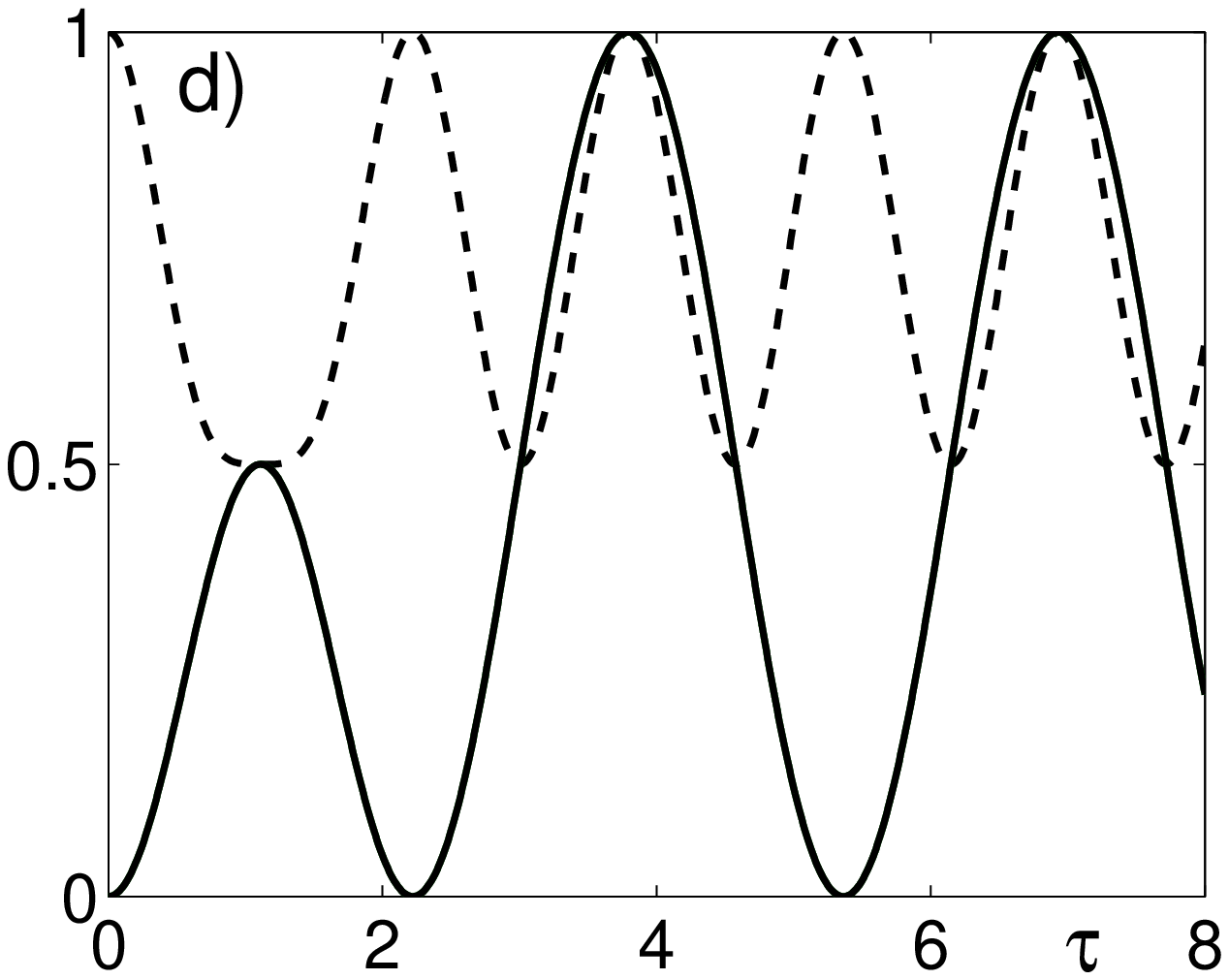}
\caption{\label{fig:fig1}Dynamics for the external field in a W
state: (a) mean photon numbers $N^{(c)}$(dashed), $N^{(f)}$(dotted)
and atomic probability $p_e$(solid); (b) tripartite negativity
$E^{(\alpha)}$ for atoms (solid), cavity modes (dashed) and external
field (dotted); ( c) atomic fidelity $F^{(a)}$ (solid) and purity
$\mu^{(a)}$ (dashed); (d) cavity mode fidelity $F^{(c)}$ (solid) and
purity $\mu^{(c)}$.}
\end{figure}

We have solved the dynamics for several types of entangled external
radiation prepared in mixed or pure states, cavity mode prepared in
the vacuum state and different types of atomic preparation. Here we
concentrate on the case of atoms in the lower energy state $\ket{ggg}_a$
and the external field prepared in a qubit-like entangled state
because this is the condition for maximum entanglement transfer to
the atomic subsystem \cite{Casagrande}. Overall we are thus dealing
with a 9-qubit system. From now on, for sake of simplicity, we take
equal coupling constants of the cavity modes with the atoms and the
external radiation ($g_c=g_a\equiv g$) and we deal with
dimensionless times $\tau\equiv gt$. We first consider the case of
negligible cavity and atomic decays. We show that it is possible to
map the initial external radiation states onto the atomic and the
cavity subsystems states, thereby also transferring the initial
quantum correlations. Due to the qubit-like form of the external and
cavity mode fields we quantify the tripartite subsystems
entanglement by the tripartite negativity $E^{(\alpha)}
(\tau)$($\alpha=a,c,f$) \cite{Sabin}, defined as the geometric mean
of the three bipartite negativities \cite{Vidal}. In
Fig.~\ref{fig:fig1} we illustrate results for the case of external
radiation prepared in the W state
$\ket{\Psi(0)}_f=(\ket{001}_f+\ket{010}_f+\ket{100}_f)/\sqrt{3}$. We
notice that each subsystem (A,B,C) has the same dynamics that is
composed by a transient and an oscillating regime. In the transient
each flying qubit transfers its excitation to the cavity qubit which
in turn passes it onto the atomic one, as shown in
Fig.~\ref{fig:fig1}a. This is the physical process for the transfer
of quantum information. The cavity mode, simultaneously coupled to
the external field and the atom (describable as a third harmonic
oscillator), exchanges energy according to a Tavis-Cummings dynamics
at an effective angular frequency $g\sqrt{2}$
\cite{Haroche_Raimond,TC}. During the transient up to time
$\tau_{off}=\pi/\sqrt{2}$ the mean photon number
$N^{(c)}(\tau)\equiv\langle \hat{c}^\dag \hat{c}\rangle(\tau)$ in
each cavity completes a cycle. In the same period the atomic
excitation probability $p_e(\tau)$ reaches its maximum value, while
the input field has completely entered the cavity, i.e. its mean
photon number $N^{(f)}(\tau)\equiv\langle \hat{f}^\dag
\hat{f}\rangle(\tau)$ vanishes.
 \\
In Fig.~\ref{fig:fig1}b we show $E^{(\alpha)}$ and see that up to
$\tau_{off}/2$ the entanglement of the external driving field is
mainly transferred to the cavity modes and then allows the build up
of atomic entanglement. We note that $E^{(a)}(\tau_{off})\cong 0.94$
which is the value of the tripartite negativity for the  injected W state. 
For the time evolution
of the atomic and cavity mode subsystems described by the density
operator $\hat{\rho}_{\alpha}(\tau)$ ($\alpha=a,c$) we show in
Figs.~\ref{fig:fig1}(c,d) the purity
$\mu^{(\alpha)}(\tau)=\hbox{Tr}_{\alpha}[\hat{\rho}^2_{\alpha}(\tau)]$
and the fidelity $F^{(\alpha)}(\tau)={}_{\alpha}
\meanvalue{\Psi(0)}{\hat{\rho}_{\alpha}(\tau)}{\Psi(0)}_{\alpha}$
with respect to a pure state $\ket{\Psi(0)}_{\alpha}$ that is a map
of the initial external field $\ket{\Psi(0)}_{f}$, with the
correspondence $\ket{0}_f\leftrightarrow\ket{g}_a$ ($\ket{0}_c$) and
$\ket{1}_f\leftrightarrow\ket{e}_a$ ($\ket{1}_c$) for the atoms
(cavity mode) states. We see that up to $\tau_{off}$ the dynamics
maps the whole initial state
$\ket{\Psi(0)}_f\otimes\ket{000}_c\otimes\ket{ggg}_a$ onto
$\ket{000}_f\otimes\ket{000}_c\otimes \hat
{U}_{-\frac{\pi}{2}}\ket{\Psi(0)}_a$, where
$\hat{U}^{(a)}_{\phi}=\bigotimes_{J}e^{-i\phi\hat{\sigma}_J^{\dag}\hat{\sigma}_{J}}$
is a local phase rotation, that in the case of W states simply acts
as a global phase factor $+i$. Note that the maximum of
$E^{(c)}(\tau_{off}/2)$ does not correspond to a pure state, i.e.
the initial state $\ket{\Psi(0)}_f$ cannot be mapped onto the cavity
modes during the transient regime.\\
\indent At the end of the transient regime the external radiation is
turned off and the subsequent dynamics can be described by a triple
Jaynes-Cummings \cite{JC} ruled by oscillations at the vacuum Rabi
frequency $2g$, hence with a dimensionless period $\pi$ as shown by
cavity mean photon number and atomic probability in
Fig.~\ref{fig:fig1}a. The purities $\mu^{(a,c)}(\tau)$ in
Figs.~\ref{fig:fig1}(c,d) oscillate at a double frequency between
pure full tripartite entangled and separable states. In particular,
at times $\tau_m=\tau_{off}+m\pi$ ($m=0,1,2...)$ the atoms are in
the entangled states $\hat{U}^{(a)}_{\phi}\ket{\Psi(0)}_a$, where
$\phi=\mp\frac{\pi}{2}$ and $-(+)$ applies in correspondence to even
(odd) values of $m$, that are the peaks of $E^{(a)}(\tau)$ in
Fig.~\ref{fig:fig1}b. At times
$\tau_n=\tau_{off}+(n+\frac{1}{2})\pi$ ($n=0,1,2...$) the cavity
modes are in the qubit-like state
$\hat{U}^{(c)}_{\phi}\ket{\Psi(0)}_c$, where
$\hat{U}^{(c)}_{\phi}=\bigotimes_{J}e^{-i\phi\hat{c}_J^{\dagger}\hat{c}_{J}}$
is a local rotation such that $\phi=0$ ($\phi=\pi$) for even (odd)
values of $n$. As regards the bipartite subsystem entanglement,
evaluated after tracing over all other qubits, we find sizeable
two-atom and two-cavity entanglement (negativity peaks $\simeq
0.41$), as well as some entanglement for all other qubit pairs
(equal
negativity peaks $\simeq 0.08$).\\
\indent Starting, instead, from an external field in the qubit-like
GHZ state
$\ket{\Psi(0)}_f=\frac{1}{\sqrt{2}}(\ket{000}_f+\ket{111}_f)$ we
find a quite similar time evolution but different values for the
maxima of the quantities evaluated in Figs.~\ref{fig:fig1}(a,b). In
particular, at times $\tau_m$ we have the maximum entanglement
transfer with $E^{(a)}=1$, that is again the same value of
$\ket{\Psi(0)}_f$, and the atomic states are mapped onto
$(\ket{000}_a\mp i\ket{111}_a)/\sqrt{2}$. On the other hand, at
times $\tau_n$ the cavity modes are in the state
$(\ket{000}_c\pm\ket{111}_c)/\sqrt{2}$. The two-qubit entanglement
is null for any pair, reflecting the lack of subsystem entanglement
in a GHZ state, except the three directly coupled atom-cavity pairs,
where the negativity peaks occur at about $0.21$ at any quarter and
three quarters of a period. We remark that the above state mapping
can be obtained also for any qubit-like state $\ket{\Psi(0)}_f$
written in a generalized Schmidt decomposition \cite{Dur}.

\subsection{Dissipative effects}
 In the perspective of experimental implementation of our
scheme for quantum information purposes an important issue is the
effect of dissipation on both state mapping and entanglement
transfer. For cavity decay rates in the range $0<\tilde{k}<0.4$ we
calculated the fidelities $F^{(\alpha)}(\tau)$ at the first peaks
(after the transient) as well as the tripartite negativities at the
first two peaks $E^{(\alpha)}$ of both atomic and cavity field
subsystems, for external field preparations in states W and GHZ. The
behavior of all these quantities can be well described as
$f(\tilde{k})=f(0)e^{-\beta \tilde{k}}$, where the values of decay
rates $\beta$ are reported in Table~\ref{tab:table1}.
\begin{table}[h!]
\caption{\label{tab:table1} Effect of cavity decay rate $\tilde{k}$
on $F^{(a)}(\tau_m)$ at the first peak ($m=0$) and $E^{(a)}(\tau_m)$
at the first two peaks ($m=0,1$), and on $F^{(c)}(\tau_n)$ ($n=0)$
and $E^{(c)}(\tau_n)$ ($n=0,1$) for injected field states W and
GHZ.}
\begin{indented}\item[]
\begin{tabular}{c|ccc|ccc}
\br
\mr
 & $F^{(a)}(\tau_{0})$ &$E^{(a)}(\tau_{0})$&$E^{(a)}(\tau_{1})$&$F^{(c)}(\tau_{0})$&$E^{(c)}(\tau_{0})$ & $E^{(c)}(\tau_{1})$ \vspace{1mm}\\
 \hline
 $\beta_{W}$ & 0.55 & 1.09 & 4.59& 1.41&3.22&6.47 \\
 $\beta_{GHZ}$ & 0.77 & 1.10&4.69&1.88&2.80&6.11\\
\br
\end{tabular}
\end{indented}
\end{table}
As expected, quantum state mapping and entanglement transfer are by
far more efficient onto atomic than cavity qubits. Actually these
processes involve the atoms when the cavity modes, coupled to the
external environment, are negligibly excited. Cavity modes are
instead affected by dissipation already in the transient regime,
which leads to a reduction of the relevant peaks as well as
asymmetries in their profile (decay slower than build up). All these
features, imprinted in the transient, become quite manifest in the
subsequent cavity dynamics.\\
\indent We also analyzed the effect of the spontaneous emission of
atomic excited levels on state mapping and entanglement transfer
processes. We consider, as an example, the case of decay rate
$\tilde{k}=0.1$ and values of dimensionless parameter
$\tilde{\gamma}$ up to 0.1.  The exponential approximation is less
accurate and we report in Table~\ref{tab:table2} the results for
exponential decay rate $\delta$ only at the first peaks of fidelity
and tripartite negativity for both atomic and cavity mode
subsystems.
\begin{table}[h!]
\caption{\label{tab:table2} Effect of atomic decay rate
$\tilde{\gamma}$ on $F^{(a)}(\tau_m)$ and $E^{(a)}(\tau_m)$ at the
first peak ($m=0$), and on $F^{(c)}(\tau_n)$  and $E^{(c)}(\tau_n)$
for $n=0$, for injected field states W and GHZ.}
\begin{indented}\item[]
\begin{tabular}{c|cc|cc}
\br \mr
 & $F^{(a)}(\tau_{0})$ &$E^{(a)}(\tau_{0})$&$F^{(c)}(\tau_{0})$&$E^{(c)}(\tau_{0})$ \vspace{1mm}\\
 \hline
 $\delta_{W}$ & 1.01 & 2.06 & 1.85& 4.07\\
 $\delta_{GHZ}$ & 1.57 & 2.21&3.14&4.51\\
\br
\end{tabular}
\end{indented}
\end{table}
\section{Fully tripartite ESD and ESB}
\begin{figure}[h]
a)\includegraphics[scale=0.42]{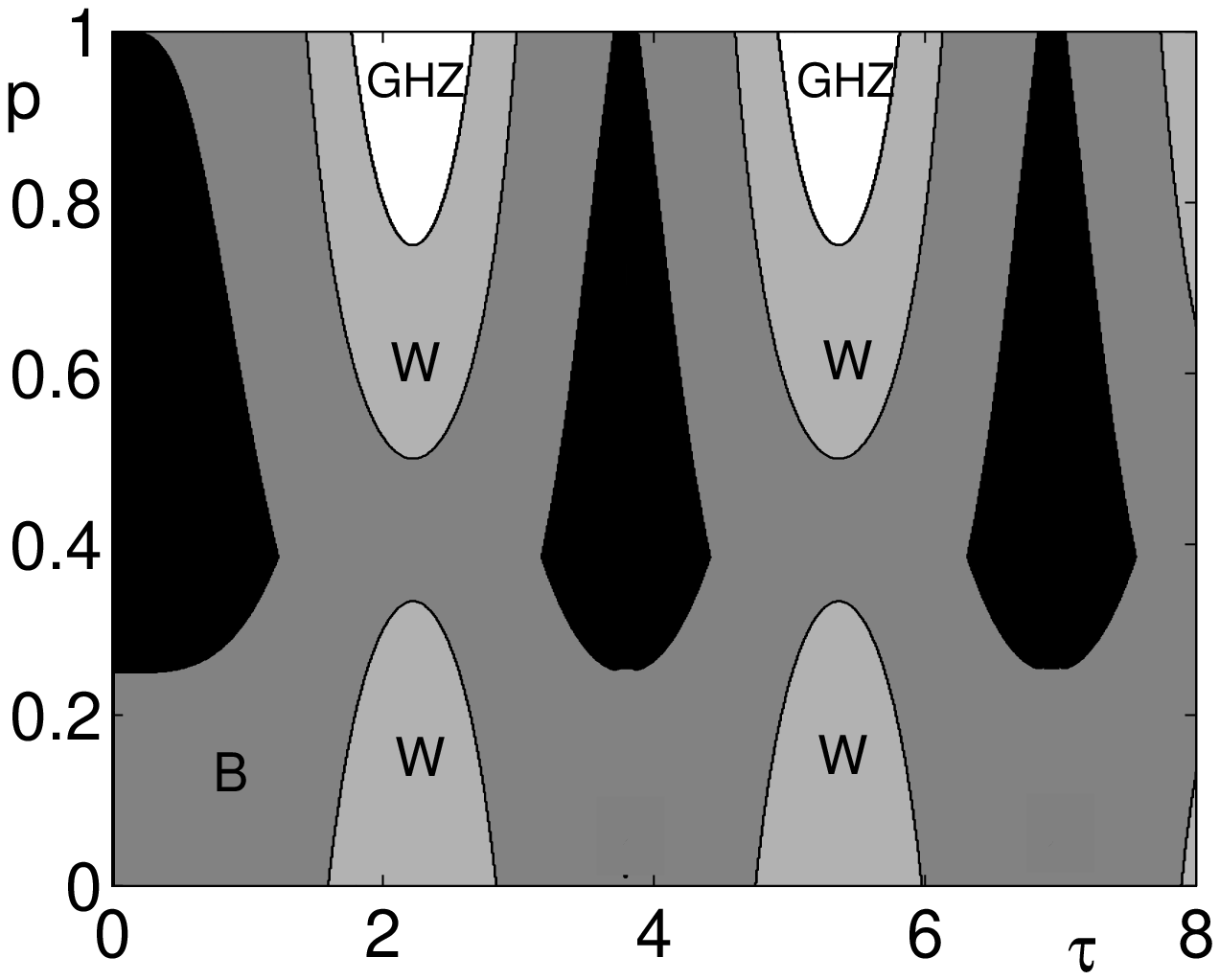}
b)\includegraphics[scale=0.42]{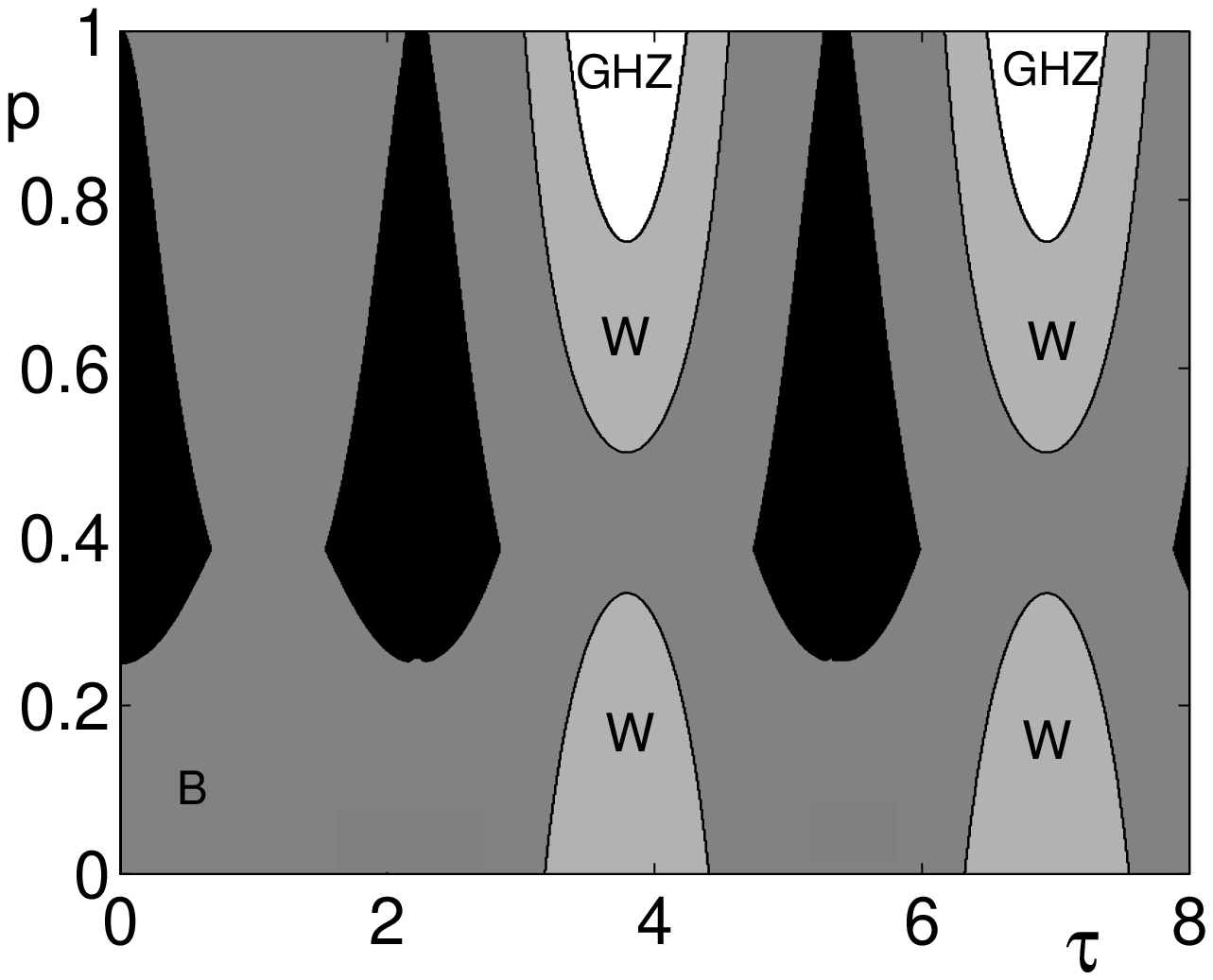}
\caption{\label{fig:fig2} Entanglement classification
 in the parameter space $\{p,\tau\}$ for external field prepared in the mixed GHZ and W state
$\hat{\rho}_f(0)$. a) atomic subsystem, b) cavity mode subsystem.}
\end{figure}

Let us now consider the class of mixed qubit-like states for the
injected field
$\hat{\rho}_f(0)=p\ketbra{GHZ}{GHZ}+(1-p)\ketbra{W}{W}$, $0\leq
p\leq 1$. By analytical and numerical results we show that our
scheme for entanglement transfer and swapping is also relevant for
the observation of sudden disentanglement and entanglement effects.
Since the tripartite negativity $E^{(\alpha)} (\tau)$ with $(\alpha=
a, c)$ is an entanglement measure that provides only a sufficient
condition for entanglement detection, we cannot properly talk about
ESD or ESB for these kind of states in the whole parameter space
$\{p,\tau\}$. Nevertheless, we can classify the atomic state by
using the entanglement witnesses \cite{Acin}
$\hat{W}_G=\frac{3}{4}\hat{I}-\ketbra{GHZ}{GHZ}$,
$\hat{W}_{W2}=\frac{1}{2}\hat{I}-\ketbra{GHZ}{GHZ}$ and
$\hat{W}_{W1}=\frac{2}{3}\hat{I}-\ketbra{W}{W}$, and analyze the
discontinuous evolution of entanglement focusing only on the
 fully tripartite entanglement properties.
For negligible dissipation and at times $\tau_m$, the initial state
of radiation is mapped onto the atoms, then the entanglement
classification is known. The state is of class GHZ for
$\frac{3}{4}\leq p \leq 1$, of class W for $0\leq p<\frac{1}{3}$ and
$\frac{1}{2}\leq p<\frac{3}{4}$, and biseparable (B) for
$\frac{1}{3}\leq p<\frac{1}{2}$. Outside times $\tau_m$, we can
still make a partial entanglement classification that is shown in
Fig.\ref{fig:fig2}a (Fig.\ref{fig:fig2}b) for atomic (cavity field)
states. Since the tripartite negativity is zero in the black regions
of Fig.\ref{fig:fig2}, we cannot exclude the presence of bound
entangled states. With this knowledge on the entanglement properties
in parameter space, we can anyway affirm that there is a
discontinuity for the full tripartite entanglement. Fixing, for
instance, a value of $p<0.25$ and looking at the time evolution of
the atomic state, we can notice that it suddenly acquires a fully
tripartite entanglement entering the W region and loses this
property after some finite time exiting that region. This effect can
be addressed as an ESD and ESB of the fully tripartite
inseparability only. Moreover, since the systems share among its
subsystems energy and entanglement in a periodic way, we can
highlight also that discontinuities in the fully tripartite
entanglement are exchanged among them. This occurs after the
transient, where the cavity modes do not exhibit genuine tripartite
entanglement (see Fig. \ref{fig:fig2}b) but only support the
transfer of quantum correlations from the input field to the atoms.

\section{Conclusions}
In this paper we have addressed the transfer of quantum information
and entanglement from three flying qubits to three localized ones,
focusing on the basic physical features characterizing multiqubit
state mapping in a CQED setting. We analyzed the effect of
dissipation at the times when the transfer protocol is optimal,
considering the atoms and the cavities both in contact with a same
reservoir at zero temperature. We derived also the conditions for
the repeated occurrence of discontinuous exchange of quantum
correlations among the multipartite subsystems. Our scheme could be
implemented by combining current advances with optical fibers,
cavities, and trapped atoms \cite{Nussmann,Serafini,Bina}.

\end{document}